\documentclass[submission,copyright,creativecommons]{eptcs}
\providecommand{\event}{FMAS 2020} 

\usepackage{underscore}           
\usepackage{graphicx}
\usepackage{epstopdf}
\usepackage{epsfig}
\usepackage{pdflscape}
\usepackage{lscape}
\usepackage[export]{adjustbox}
\usepackage{colortbl}
\usepackage{amsmath}
\usepackage{amsfonts}
\usepackage{amssymb}
\usepackage{setspace}
\usepackage{bookman}
\usepackage{charter}
\usepackage{mathpazo}
\usepackage{mathptmx}
\usepackage{verbatim}
\usepackage{moreverb}
\usepackage{listings}
\usepackage{etoolbox}
\patchcmd\section{2.3ex}{1ex}{}{}
\patchcmd\subsection{1.5ex}{.9ex}{}{}
\patchcmd\subsubsection{1.5ex}{.8ex}{}{}
\title{A Formal Model for Quality-Driven Decision Making in Self-Adaptive Systems}
\author{Fatma Kachi \qquad Chafia Bouanaka\qquad Souheir Merkouche
\institute{LIRE Laboratory\\University of Constantine2-Abdelhamid Mehri\\
Constantine, Algeria}
\email{(fatma.kachi, chafia.bouanaka, souheir.merkouche)@univ-constantine2.dz}
}
\def\titlerunning{A Formal Model for Quality-Driven Decision Making in Self-Adaptive Systems}
\def\authorrunning{F. Kachi, C. Bouanaka \& S. Merkouche}
\begin{document}
\maketitle

\begin{abstract}
Maintaining an acceptable level of quality of service in modern complex systems is challenging, particularly in the presence of various forms of uncertainty caused by changing execution context, unpredicted events, etc. Although self-adaptability is a well-established approach for modelling such systems, and thus enabling them to achieve functional and/or quality of service objectives by autonomously modifying their behavior at runtime, guaranteeing a continuous satisfaction of quality objectives is still challenging and needs a rigorous definition and analysis of system behavioral properties. Formal methods constitute a promising and effective solution in this direction in order to rigorously specify mathematical models of a software system and to analyze its behavior. They are also largely adopted to analyze and provide guarantees on the required functional/non-functional properties of self-adaptive systems. Therefore, we introduce a formal model for quality-driven self-adaptive systems under uncertainty. We combine high-level Petri nets and plausible Petri nets in order to model complex data structures enabling system quality attributes quantification and to improve the decision-making process through selecting the most plausible plans with regard to the system's actual context.  

\noindent\textbf{Keywords:} Formal methods, Petri nets, Self-adaptive systems, Quality-driven systems, uncertainty models.
\end{abstract}
\section{Introduction}
Modern advanced software systems are required to perceive important structural and dynamic changes to their operational environment as well as to their internal status, and to adapt to these continuous changes autonomously ~\cite{DeLemos2013}. They aim to achieve better quality of service and ensure the required functionality. However, such systems are expected to deal seamlessly with different types of uncertainty during operation. These uncertainties are often difficult to predict at design time, requiring software to be deployed with incomplete knowledge and handle changing conditions during operation ~\cite{Weyns2017}. Consequently, software engineers are investigating new techniques to handle uncertainty at runtime without incurring penalties, which is commonly referred to as self-adaptation ~\cite{DeLemos2013, DannyWeyns2017}. Many software systems actually need to comply with strict requirements, providing guarantees for system properties such as ensuring a certain level of performance and reliability.

Self-adaptability ~\cite{Laddaga2004} is a well-established approach for modelling such systems, and allowing them to be able to achieve functional and/or quality of service objectives by autonomously modifying their behavior at runtime. The MAPE-K ~\cite{IBM2006, Arcaini2015}  (Monitor-Analyze-Plan-Execute over a shared Knowledge) loop is one popular approach that has proven its efficiency in modelling self-adaptability since it covers the necessary activities to be performed in a control loop. However, uncertainty is a fundamental challenge of SASs (Self-Adaptive Systems). It involves not only system requirements but also its execution context and affects the system quality of service. Therefore, self-adaptation mechanisms driven by quality modify system behavior dynamically. Albeit there has been extensive research to address uncertainty in SASs ~\cite{Shevtsov2019, DeLemos2017, Mahdavi-Hezavehi2017}, there is no focus on proposing solutions to identify uncertainty at different levels of the decision-making process and considering it when modelling the SASs. Besides, the main focus has been on achieving adaptations without determining their side effects on the overall system qualities. Moreover, existing approaches do not allow choosing the most appropriate adaptation plan in terms of the best side effects.

The engineering of quality-driven self-adaptive systems, evolving under uncertainty, needs to consider the above issues. Besides, designing this type of systems requires a verification and validation phase using formal methods and tools so that the model can be analyzed with regard to quality. Formal methods constitute a promising and effective solution to rigorously specify mathematical models of a software system and analyze its behavior. They are also largely adopted to analyze and provide guarantees on the required properties of self-adaptive systems. In this field, Petri nets ~\cite{Wang2007} have shown their ability as a powerful tool for modelling and verifying complex systems, a number of variants have also been introduced to help modelling uncertainties, such as fuzzy Petri nets ~\cite{Looney1988, Ding2018}, possibilistic Petri nets ~\cite{Lee2003}, etc.

In this paper, we introduce a formal model for self-adaptive systems evolving in dynamic environments and execution contexts. We mainly leverage a MAPE-K loop-based model, combining high-level Petri nets (HLPNs) and plausible Petri nets (PPNs), for the design of quality driven systems under uncertainty. Although PPNs allow managing uncertainty and guiding the decision-making process, they do not support the complex data structures that are necessary to enable the quantification of system quality properties. To this end, we exploit the expressive power of HLPNs. We mainly extend the model proposed in ~\cite{Camilli2018} to capture uncertainty by combining PPNs with the already used HLPNs. The basic approach does not consider the uncertainty concerns in the decision-making process. It does not also allow representing complex information and characteristics of a dynamic system owing to the use of ordinary Petri nets. In addition, the separation of concerns between system qualities may lead to negative side effects. Therefore, we use HLPNs to represent complex systems and data structures; and PPNs to assist the decision-making phase in selecting the best plans in the presence of uncertainty. 

The remainder of this paper is structured as follows. Section 2 recalls basic concepts on Petri net types involved in our model. Section 3 discusses related work on modelling SASs and tackling the problem of uncertainty. Section 4 formally describes the proposed model and its components, the latter is illustrated and validated through the problem of the aircraft planning in Section 5. Finally, Section 6 concludes the paper.
\section{Background}
Petri nets have been initially proposed to model the behavior of a dynamic system with discrete events ~\cite{Petri1962}; they have then undergone several evolutions and variants  ~\cite{InternationalStandardISO/IEC159092000, Chiachio2019, Baez2020, Taleb-Berrouane2020} to cover more concerns in system modelling and analysis. In what follows, we present two types of Petri nets to be used in this work, high-level Petri (HLPNs) nets and plausible Petri nets (PPNs). To facilitate the understanding of our approach and the usefulness of Petri nets types to be adopted, we give only an informal description of what they are and their purpose; we refer the readers to  ~\cite{InternationalStandardISO/IEC159092000, Chiachio2018} for formal definitions and more details.
\subsection{High-Level Petri Nets (HLPNs)}
HLPNs are a well-defined semi-graphical technique for the specification, design and analysis of systems. HLPNs are applicable to a wide variety of concurrent discrete event systems and in particular distributed ones. Generic fields of HLPN application domains include: requirements analysis; development of specifications; modelling business and software processes; simulation of systems to increase confidence; formal analysis of the behavior of critical systems; development of Petri net support tools, etc. ~\cite{InternationalStandardISO/IEC159092000}.

An HLPN is made up on a set of nodes (i.e. places and transitions) and a set of arcs connecting places to transitions and vice versa as in an ordinary Petri net but is extended in the following way:  each place is associated with a place type and can contain a collection of tokens corresponding to that place type. Transitions are dotted with boolean expressions (for example, $ x <y $) called guards. Additionally, arcs are inscribed with expressions called arc annotations; expressions may contain constants, variables, and function images. An expression is evaluated by assigning values to each of its variables. Whenever an expression evaluates to true, a multiset of tokens is produced in the output places of the corresponding transition according to arc’s weights and types.
\subsection{Plausible Petri Nets (PPNs)}
The combination of the principles of Petri nets with the foundations of information theory resulted in a new model for Petri nets, called plausible Petri nets (PPNs) ~\cite{Chiachio2017, Chiachio2018}; which are a hybrid variant of Petri nets composed of two types of places and transitions, namely, symbolic and numerical, in order to describe both discrete and continuous behaviors of a system. In the symbolic subnet, the discrete  behavior is described using regular tokens, while in the numerical subnet, continuous or numerical behavior is described with tokens that carry information about the states of variables. A state of information about a given variable   is the probability density function (PDF) of \textit{x} over $\chi$ ~\cite{Rus2016}, where $\chi$ is the state space of a stochastic variable \textit{x}. For a numerical transition, it can fire when the conjunction between all its input places’ states of information and the transition’s states of information is possible. For a mixed transition, both conditions of symbolic and numerical transitions must be satisfied. Firing transition’s effect for the numerical places is a state of information consisting of a disjunction of the previous state of information, and the information produced after firing the transition (conjunction of state of information within the transition and its input places). For more details on PPNs, the reader is referred to ~\cite{Chiachio2019}.

The main feature of PPNs resides in their efficiency to jointly consider the evolution of a discrete event system together with uncertain information about the system state using states of information ~\cite{Chiachio2018}. They provide a mapping between the possible numerical values of a state variable and their relative plausibility, hence giving greater versatility for representing uncertain knowledge using a more principled approach ~\cite{Chiachio2019}.

\section{Related work}
Over the past years, researchers have developed a large body of work to formally model self-adaptive software systems and many approaches have shown remarkable progress in providing solutions to mitigating uncertainties in these systems using various formal approaches.

In ~\cite{Camilli2018}, the authors proposed a formal framework, based on HLPNs, to model a distributed self-adaptive system. In particular, the framework shows how the most significant concepts related to self-adaptation can be formally specified in terms of HLPN, and how structural changes implemented by multiple control loops can be described in a natural way. A two-layered architecture is adopted; it ensures a clear separation between the managed and the managing systems. The HLPN emulator is capable of representing the system dynamics; for this purpose, the basic Petri net is encoded in the emulator marking. The managing system is defined by a collection of MAPE-K control loops, specified using HLPNs. To implement sensors and actuators, the read and write API primitives are used. The approach is consistent and based on well-established formal methods. As such, it takes advantage of consolidated analytical techniques. However, a major drawback of this approach is that it does not consider uncertainties to be faced in the decision-making phase. This leads to poor system performance. Hence when choosing an adaptation plan, the system lacks global knowledge changes to be performed and therefore is unable to determine the negative effect of each plan in order to choose the best one. Besides, it does not allow modelling complex systems due to the use of ordinary Petri nets for modelling the managed systems.   

In ~\cite{Shevtsov2019}, an approach called Simplex Control Adaptation (SimCA*) is presented, it allows building self-adaptive software systems that satisfy multiple STO-reqs – a combination of S-reqs (stakeholder requirements), T-reqs (threshold requirements), and O-reqs (optimization requirements) in the presence of different types of uncertainty. SimCA* has dedicated specific components to monitor changes in the underlying system or its environment and adjust the adaptation logic accordingly to deal with different types of uncertainty. The main contribution of SimCA* is in applying formal techniques to adapt the behavior of software systems, which is one key approach for providing correctness guarantees. Formal analysis in SimCA* is based on an equation-based model of the software system and leverages on guarantees provided by basic SimCA ~\cite{Shevtsov2017}. This analysis is complemented by an empirical evaluation that demonstrates that SimCA* achieves the required quality goals. Although this approach deals with uncertainty and offers correctness guarantees, it does not support systems and uncertainties modelling.

In ~\cite{Chiachio2019}, the authors provide a framework for modelling self-adaptive expert systems (SAeSs) using Petri nets. The Petri nets used here are called plausible Petri nets combined with Bayesian learning principles. PPNs model uncertainty through information states, which provide a map between the possible numerical values of a state variable and their relative plausibility. This methodology is primarily used in SASs that deal with uncertainty, for monitoring system infrastructure assets. This model allows systems to handle uncertainties and to adapt to their occurrence; applying Bayesian learning particularly allows generating new adaptation decisions, which is a suitable solution for the adaptation problem. However, since it uses symbolic subnet which accounts for the discrete behavior of the system using regular tokens, as for ordinary Petri nets, it does not allow representing complex structured data and thus complex systems. 

In ~\cite{Ding2016}, a new type of Petri net based on neural networks to model adaptive software systems was presented. It is an extension of hybrid Petri nets by embedding a neural network algorithm into them on particular transitions; system adaptation is realized through the learning ability of neural networks. The proposed model considers the runtime environments and ensures that components collaborate to make the suitable adaption decisions while the computing happens locally. We highlight that the model is more efficient than traditional optimization solutions since it is able to not only process runtime data and make decisions, but also model the behavior of software systems. A major drawback of the proposed model is that it does not allow new decisions to be made, i.e. adaptation actions have to be defined statically, in addition to the conditions for their selection, which is a difficult and costly task in the case of complex systems.

Although some of the models presented in this section inspired us to define our model, none of them allow both modelling self-adaptation and managing uncertainty at the same time. So, we can conclude that:
\begin{itemize}
	\item Models dedicated to self-adaptive systems modelling do not generally allow the generation of new decisions, but rather the decisions are encoded beforehand and the system deduces which one to apply according to the actual context changes. Thus, when new events occur, the system does not know what to do. Moreover, the system's external environment is generally discarded, since in most cases only system resources are considered as its execution context. 
	\item Models dedicated to managing uncertainty are in most cases hybrid Petri nets that support the modelling of discrete and continuous events affecting the system, they are usually specific to a particular type of systems, giving rise to difficulties in their reuse and do not support the modelling of more complex systems. 
\end{itemize}
\section{Proposed model}
With the emergence of complex systems, ensuring their effectiveness and efficiency is very difficult due to variations in their execution contexts and evolution in their requirements. Through self-adaptation, a system is able to cope with these contextual and environmental changes, and hence adapting to the new conditions in which it evolves. However, in the case of a quality-driven self-adaptive system, the major difficulty lies in selecting the most appropriate adaptation plans, adaptation actions and side effects that ensure and maintain the required system global qualities. Consequently, one major challenge is to deal with and to be able to exploit such uncertainty to dynamically adapt the modelled system.
In this paper, we aim to combine HLPNs and PPNs to model quality-driven systems under uncertainty while considering the necessary artifacts to quantify system qualities and guide the decision-making to select the best plans with regard to quality attributes. 
\subsection{Overall architecture}
\begin{figure}
	\centering
	\includegraphics [scale = 0.5] {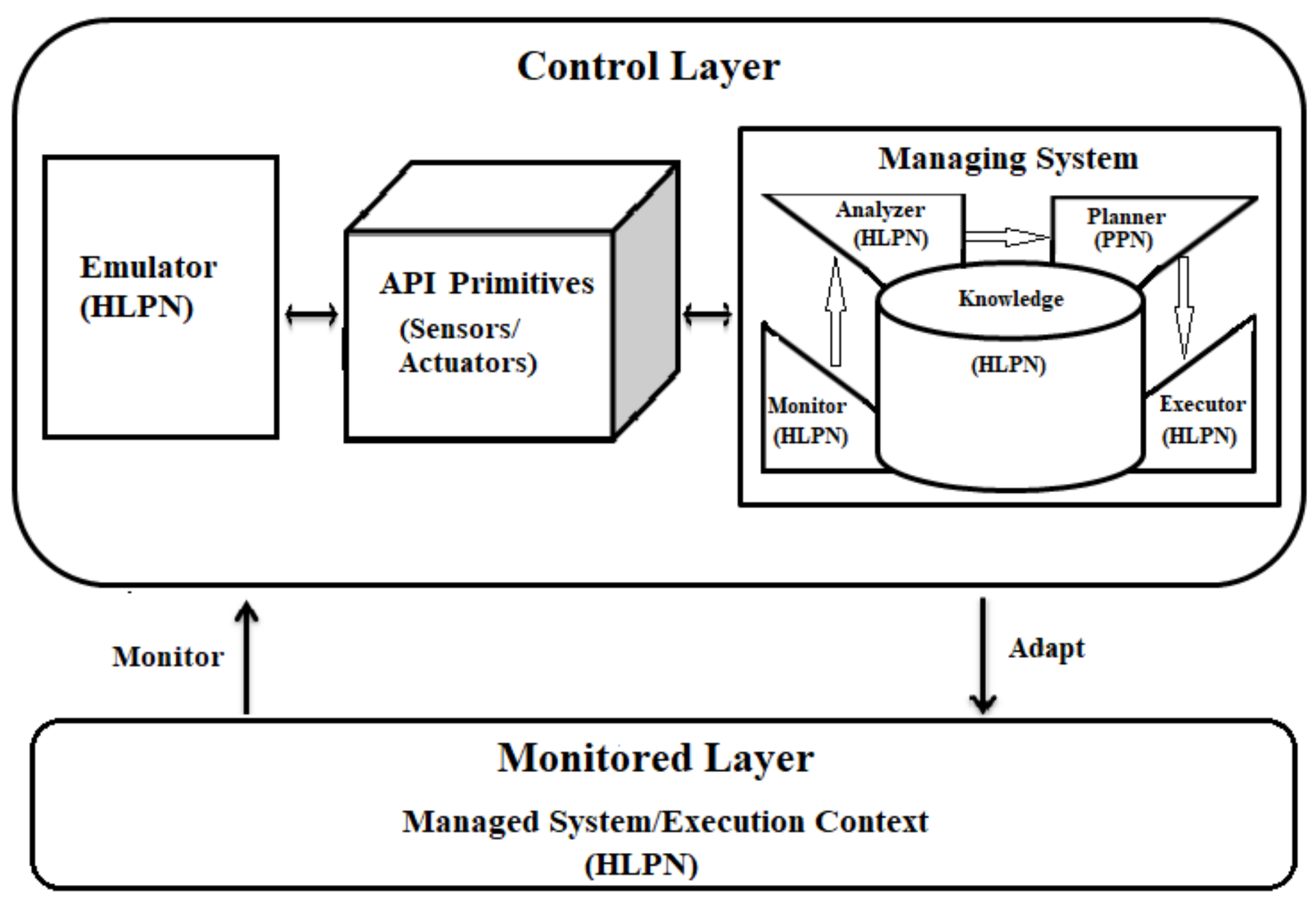}
	\caption{ Approach Overview.}
	\label{fig: archi}
\end{figure}
In ~\cite{Camilli2018}, HLPNs were adopted to formally model distributed self-adaptive systems. Although the two-layered architecture defined in ~\cite{Camilli2018} ensures a clear separation between the managed and the managing systems, its major drawback is that it does not consider the uncertainty concerns in the decision-making process. It also does not allow representing complex information and characteristics of a dynamic system owing to the use of ordinary Petri nets, which manipulate a single type of tokens and thus lack the necessary expressive power to model the managed system and its execution context. In addition, the managing system is defined by a collection of MAPE-K control loops, specified by means of HLPNs, in order to define a loop for each adaptation concern or quality. Such separation of concerns between system qualities may lead to negative side effects i.e. when the system tries to improve one quality, it may negatively alter other ones. 

In our approach, we adapt and extend the model proposed in ~\cite{Camilli2018} to formally model quality-driven systems under uncertainty and to maintain several qualities. Our objectives through this extension are:
\begin{itemize}
	\item to mitigate uncertainties in order to ensure the continuous satisfaction of system qualities as effectiveness, efficiency, reliability...
	\item to assist the decision-making process while selecting the proper adaptation plans in order to maintain the desired quality of service. 
	\item to improve the model expressiveness and allow representing more complex information of a dynamic system; the managed system and its environment for instance.
\end{itemize}
To achieve these objectives and be able to model a quality-driven SAS, we need to firstly identify the overall qualities of the system, and then determine the system characteristics that allow their quantification as well as the contextual uncertainties that affect the system behavior. System qualities may include safety, which informally require that something bad will never happen, and efficiency, which means that the system will select the most efficient adaptation plans. We will explain these qualities in more detail in section 5. To this end, we combine HLPNs and PPNs; where HLPNs intervene in the definition of data flows through expressions and annotations; this feature will be exploited to quantify the observed qualities among the different elements of the model. On the other hand, PPNs are used to assist and improve the decision-making process and hence determine the most appropriate plans; we mainly exploit the plausibility concept of PPNs to select the most appropriate plan for an adaptation condition. We point out that, contrary to the approach adopted in ~\cite{Camilli2018}, we model the managing system through a single control loop in order to predict and treat negative side effects of the adaptation plans on the system global qualities. Figure ~\ref{fig: archi} depicts an overview of the proposed model which is an extension of the basic structure of the architecture proposed in ~\cite{Camilli2018} in the following directions: (1) the monitored layer is modelled by a HLPN rather than an ordinary Petri net and (2) for the control layer, the emulator is extended to support HLPNs that are also used to represent the MAPE-K elements except the planner (P) which is represented by a PPN. This model provides a rigorous means to specify systems, ensure the continuous satisfaction of their quality and perform adaptations using the most appropriate plans.	
\subsection{Monitored layer}
It encloses the definition of the managed system and its operating environment. We adopt HLPNs to model the monitored layer, it describes the global behavior where places model the managed system states or the context elements; transitions model actions to be performed by the system; tokens hold information about the element being modelled in that place and accept any data structure, which is the major benefit of HLPN in addition to the concept of expressions that allows quantifying system qualities.
\subsection{Control layer}
\begin{figure}
	\centering
	\includegraphics [scale = 0.42]{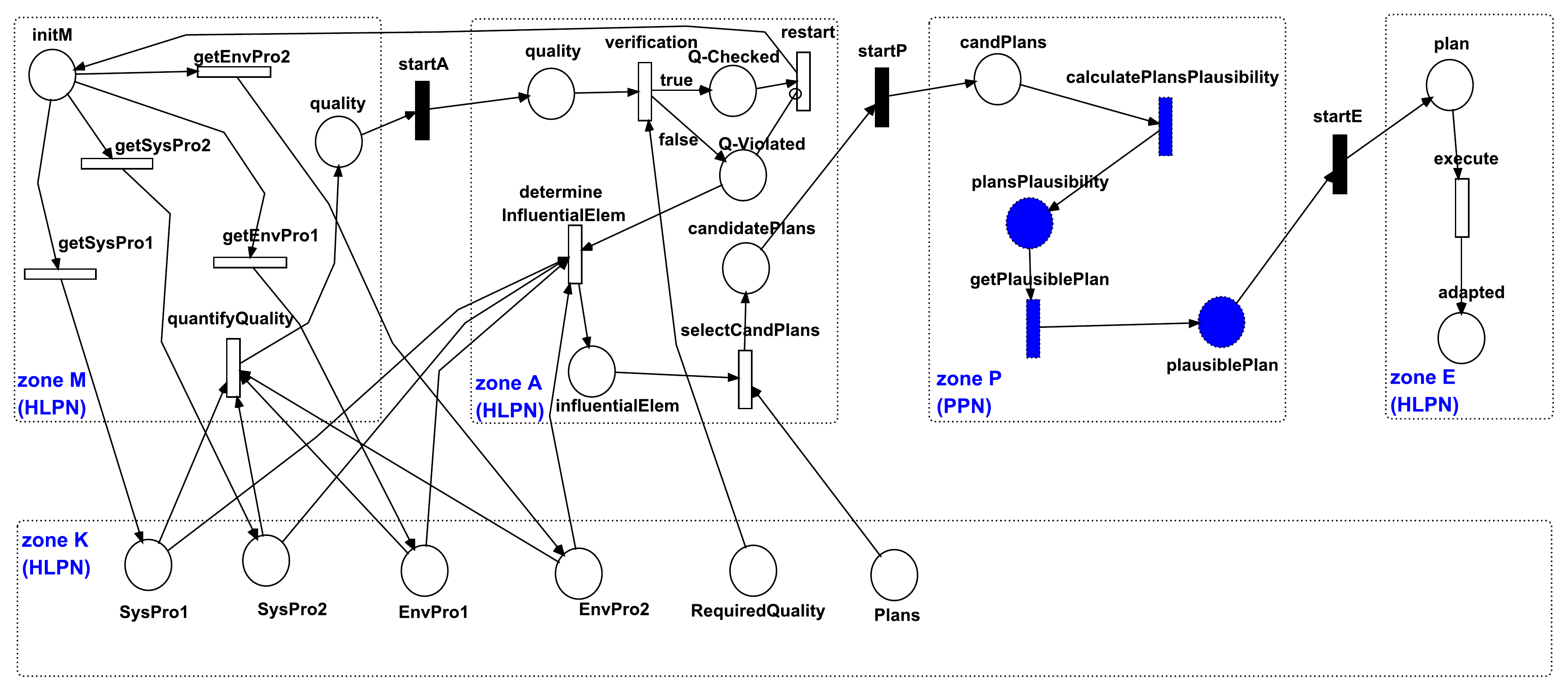}
	\caption{ Detailed View of MAPE-K loop.}
	\label{fig: ming}
\end{figure}

The control layer is generic and parameterized by the monitored layer, the quality objectives and adaptation actions. It contains (see Figure ~\ref{fig: ming}) the emulator, the managing system which is a combination of HLPN and PPN and the API consisting of a set of read/write primitives represented by HLPN transitions. These components provide specific roles to achieve a clear separation of concerns. To take charge of and be able to manipulate the HLPNs concepts, we have also extended and updated the API and the emulator elements; we give in what follows a brief description of the use of the emulator and the API elements (for the full description see ~\cite{Camilli2018}).

The emulator is used to encode the monitored layer in a specific structure modelled by an HLPN to be manipulated by firing transitions connected to places. In the emulator, each place gathers a set of elements of the managed system which are: places of the managed system together with their markings, its transitions with their guards, the input arcs with their expressions, and the output arcs with their expressions; so it has 4 places. This representation allows and facilitates the use of the API; executing the API primitives invokes the firing of the monitored layer transitions and/or adaptation of its structure. In the emulator model, the single transition move, whenever it fires, triggers the firing of a transition in the monitored layer being emulated. The emulator is connected to the managing system by means of an output arc from the transition \textit{move} to the place \textit{initM} of zone M in Figure ~\ref{fig: ming}.

The API is a set of primitives that allow reading as well as modifying the Petri nets of the monitored layer. The primitives are used inside the MAPE-K control loop to simulate the sensing and actuating actions upon the monitored layer. Each primitive is formalized by a HLPN transition (connected to specific places of the emulator) which reads or modifies the encoded Petri nets associated with the managed system in a consistent and atomic manner. 

The managing system is a MAPE-K control loop specified in terms of a combination of HLPNs and PPNs sharing the same knowledge where the quantification of the observed qualities is carried out using the data flow defined by the HLPNs and the decision-making phase or the planner element is extended to assist it in selecting the best plans in the presence of uncertainty using PPNs. An overview of the managing system detailed structure is given in Figure ~\ref{fig: ming} where blue places and transitions are plausible and the black transitions are the cross-zone transitions; it will be detailed in the following sub-sections.
\subsubsection{Knowledge}
Since the managed system is subject to and impacted by changes in the internal/external contexts, it is inevitable to monitor them to maintain the required quality. Therefore, these contexts are represented as data hold in places of zone K (see Figure ~\ref{fig: ming}). Places \textit{$SysPro_1$}, \textit{$SysPro_2$} define system properties, places \textit{$EnvPro_1$}, \textit{$EnvPro_2$} define the current environmental properties; this data is collected and updated by the monitor element. Updates in a contextual element may provoke violation of some system quality. The contextual element is thus considered to be an influential element and will be used to determine the adaptation actions to be performed. The \textit{Plans} place contains all possible adaptation  actions that are necessary to maintain the system qualities. These actions are rules of the form condition/action. 

\subsubsection{Monitor}
It is responsible for collecting the monitored data from the internal and external contexts of the managed system by executing transitions \textit{$getSysPro_1$} (\textit{$getSysPro_2$}, ...) for the system internal context and \textit{$getEnvPro_1$} (\textit{$getEnvPro_2$}, ...) for the system external context; these transitions refer to the read primitive of the API.  Then, it quantifies system quality based on the characteristics of the underlying system represented by the HLPN concepts, arcs weights and place markings; the result which is a multi-set of tokens, each one representing a system property together with its actual value, is saved in the quality place. 

\subsubsection{Analyzer}
The Analyzer element is in charge of analyzing system qualities, it represents zone A of Figure ~\ref{fig: ming} and is triggered by the firing of the cross-zone transition that removes tokens from the source zone and puts them into the target zone, the \textit{startA} cross-zone transition removes tokens of the \textit{quality} place from zone M and puts them into zone A in the \textit{quality} place. The transition called \textit{verification} has as input the actual quality value and the required quality (the quality threshold). Depending on the comparison of the two values, its firing determines whether an adaptation is required or not. Whenever an adaptation is required, a step for determining the influential element is done by firing the \textit{determineInfluentialElem} transition; since the system quality is affected by contextual elements and quality requirements. The result of this step is used as an input to select the adaptation plans that may restore the system quality. Concretely, the \textit{selectCandPlans} transition allows selecting the candidate plans to carry out an adaptation according to the actually identified influential element, which might appear in the transition guard. At the end of the analysis phase, the analyzer triggers the planner and transmits the candidate plans to the planner using the cross-zone transition \textit{startP}. 
\subsubsection{Planner }
Its task is to select the proper adaptation plan from a set of candidate plans with regard to quality requirements. The Planner element is defined via a PPN which models the decision-making process and facilitates the selection of the most plausible plan to be performed in order to both restore the violated quality attribute and maintain the rest of system qualities, i.e., avoid negative side effects. The plausibility of each candidate plan is calculated, via the \textit{calculatePlansPlausibility} transition of Figure ~\ref{fig: ming}, on the basis of the managed system data and the considered plan side effects on the other properties and the overall quality of the system. The most plausible plan is finally selected and transmitted via the \textit{plausiblePlan} place to the executor element.
\subsubsection{Executor } 
It executes the selected adaptation plan through the \textit{execute} transition of Figure ~\ref{fig: ming}, representing the write primitive of the API. In fact, the managed system structure (places and transitions of the HLPN associated to the managed system) could not be changed but are rather the system characteristics, i.e., information contained in the tokens and the weight of the arcs.

The managing system cycle iterates periodically to ensure the continuous satisfaction of the system overall qualities. The various concepts introduced by the proposed model will be illustrated via the aircraft planning case study in the next section.
\section{Case Study: Aircraft Arrival Planning}
In order to clarify the proposed model and examine its effectiveness and efficiency, we consider the problem of aircraft planning, we first describe the problem and, identify and quantify its qualities. Then, we detail its modelling.
\subsection{Problem Description}
Planning and managing air operations is becoming increasingly complicated because of their congestion, but it is essential to optimize airport capacity. In order to ensure continuous traffic on the runways and to maximize the use of the airport infrastructure, a minimum level of queuing and optimal planning is required. However, the dynamics of delays and their propagation are essential elements when assessing the performance of airports ~\cite{Cook2015}.

Planning consists of assigning each aircraft a runway, a gateway, a pair of a terminal and a gate; consisting of the parking zone of an aircraft, to complete its departure/landing procedure. This task is usually performed as desired on the day of landing, taking into consideration two main characteristics: capacity optimization and arrival/departure safety within the airport. However, a significant number of unexpected events can occur and alter this planning and thus require its update, we can cite:

\begin{itemize}
	\item An aircraft can arrive either later or earlier than planned. As a result, the resources allocated to it may be unavailable. The problem in this case is: should it wait until the resource is released, or should it be assigned to another resource? In this case, another question arises: which is the most efficient solution? And what are consequences on the rest of the planning, i.e., the other aircraft?  
	\item Change of wind direction also impacts the assigned runways since an aircraft has to land with the wind direction. If the wind direction suddenly changes, would there be enough time to redo the landing plan for a set of aircraft without causing delays?
	\item Other events can also occur suddenly affecting the planning, such as a resource break down, or an aircraft occupying a resource longer than was estimated, adverse weather conditions, etc. 
\end{itemize}
These events lead the planning system to behave in an uncertain manner; our objective throughout this paper is to manage these uncertainties by defining a model that is capable of adapting and updating the aircraft planning in the presence of a changing environment or context, but still maintaining safety and efficiency constraints. For the air operations, safety concerns maintaining a minimum separation between two consecutive aircraft, i.e., It is necessary to ensure that a certain distance always separates two consecutive aircraft according to their types. The efficiency of the air operations is achieved by minimizing delays. Efficiency consists of choosing an available resource that reduces or prevent the delays. Although these constraints are statically checked during the initial planning, they need to be verified again due to alterations and updates of the effective planning.

In this case study, we are interested in the aircraft arrival management system and its subsequent problems. The problem of arrival sequencing and scheduling at a given destination airport has been studied for several decades ~\cite{1978, Bennell2011, Bennell2013}. The aircraft is indeed moving in a constrained and potentially congested space. The arrival procedure consists of five phases: approach is the starting point of the procedure, i.e. the aircraft enters the airport zone; sequence consists of assigning a sequence number to the aircraft, as soon as it gets confirmation that the runway is clear, the aircraft flies to that runway; landing means that the aircraft has arrived at the runway and is landing; taxiing consists of rolling  the aircraft to the terminal on its assigned track; and parking is the final phase; the aircraft has arrived at the gate where it is programmed to park.

\subsection{A self-adaptive structure for the Aircraft planning system}
Our proposed model is general and can be applied in several areas and on different cases. However, in order to be able to apply it to a given case study, we first project it on that case study and identify the necessary parameters of the control layer.
\subsubsection{Modelling the aircraft arrival procedure }
\begin{figure}
	\centering
	\includegraphics [scale = 0.42]{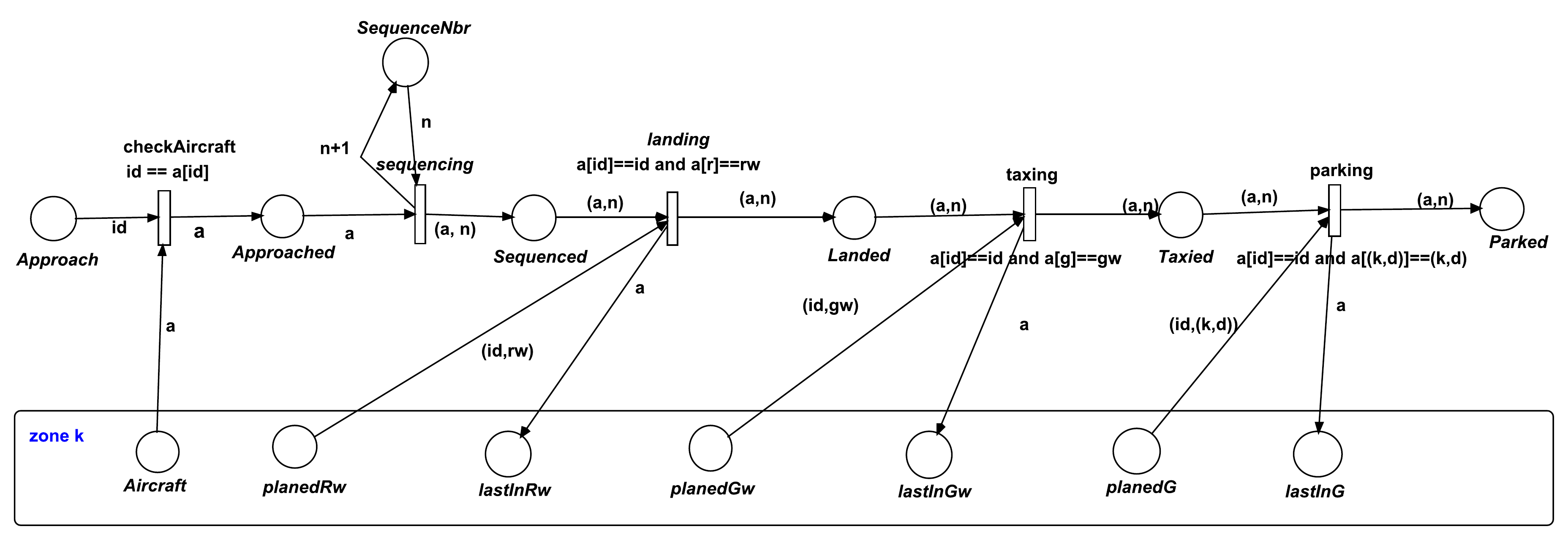}
	
	\caption{ Formal modelling of the aircraft arrival procedure.}
	\label{fig: med}
\end{figure}

Figure ~\ref{fig: med}  represents the monitored layer model of aircraft’s arrival procedure. The places \textit{Approached}, \textit{Sequenced}, \textit{Landed}, \textit{Taxied} and \textit{Parked} represent the different phases of the procedure, the presence of a token in a given place means that the aircraft has successfully finished the corresponding phase. Aircraft being at the approach phase are passed by a step of checking that they are among those planned, this is illustrated by the \textit{checkAircraft} transition; the planning system retrieves the aircraft information and puts it in the \textit{approached} place and assigns it a sequence number in the \textit{sequenceNbr} place. Places \textit{Sequenced}, \textit{Landed} and \textit{Taxied} constitute the inputs of the MAPE-K loop to check runways, gateways and gate availability and constraints, respectively. Places \textit{planedRw}, \textit{planedGw} and \textit{planedG} represent outputs of the MAPE-K loop; tokens in these places allow the aircraft to realize its activity. Transitions represent the aircraft movement from one phase to another. In this case study, we consider the following data: an aircraft is identified by a tuple ($ id, c, r, g, (k, d), ts, t, tr, tg, tp, tk, tf $) where $ id $ is the aircraft identifier; $ c $ represents the aircraft category (Large (H), Medium (M), Light (L)); $ r $, $ g $, and $ (k, d) $ represent a runway, a gateway, and a pair of terminal and gate, respectively; $ ts $ is the time of the aircraft occurrence at the sequence point; $ t $ is the estimated aviation time to the landing point; $ tr $ is the planned landing time of the aircraft, i.e., the entrance to the planned runway; tg is the planned time to enter the planned gateway; tp is the estimated taxiing time to arrive at the parking point; $ tk $ is the arrival time planned to reach the gate; $ tf $ is the planned exit time of the gate. A runway is identified by ($ r, rs, er $) where $ r $ is the runway identifier; $ rs $ is the runway state (free, occupied, or inoperative); er is the identifier of the emergency runway of the actual runway. Generally, when a runway is used, the opposite runway (in direction) is free because of wind constraint. Each gateway is characterized by ($ g, gs $) where $ g $ is a gateway identifier; $ gs $ is the gateway state (free, occupied, or inoperative).
\subsubsection{Identifying air operations qualities and constraints}
The main objective of adapting the aircraft planning is to ensure both safety and efficiency requirements. Safety concerns maintaining a minimum separation distance between two consecutive aircraft and mainly avoiding collisions problems. The efficiency constraint refers to reducing delays while assigning resources to aircraft. Therefore, the aircraft arrival planning system aims to define landing plans that ensure the safety of passengers while avoiding delays through selecting the most effective ones.
\\

\noindent\textbf{a) Securing the arrival procedure (safety)}\\
Safety concerns maintaining a minimum separation distance between two successive aircraft at various points in the arrival procedure according to their categories ~\cite{Skorupski2016}. In our work, it has been set at one-minute separation for cases not listed in ~\cite{Skorupski2016}; the system administrator may still update these parameters. The separation constraint is represented by the place Separation. The analyzer is responsible for checking the arrival procedure safety at various points and then sends the results to the planner; separation constraints are resumed in Table \ref{tab: sep} below. 
\\
\textbf{b) Planning effectiveness}\\
The effectiveness constraint is generally met by allocating new resources to the delayed/affected aircraft, but still maintaining the safety constraint and reducing the delay time. Whenever an aircraft is late, it is obvious that it will complete its procedure after the estimated time causing alterations to the successor aircraft planning and resource occupancy. In this case, it will be necessary to try to find other resources for the successor aircraft to complete its procedure on time. An aircraft may also arrive early, and hence the resources may be not yet available. In this case, new resources have to be found to avoid and reduce its waiting time. A simple measure of the planning effectiveness will depend on the safety and serviceability of the aircraft.

The wind direction metric is also considered; whenever the wind direction changes, aircraft must be reassigned to other runways. Since the opposite runway of a planned one is always unoccupied, the solution consists of switching the landings to the opposite runways while maintaining their order.

\begin{table}
	
	\caption{Aircraft arrival separation constraints.}
	\begin{tabular}{|p{1.2cm}|p{5.15cm}|p{8.2cm}|}
		\hline \textbf{Phase} & \textbf{Separation constraints}& \textbf{Description}  \\
		\hline Landing & $ S (ai, ai+1) > tr(ai+1) – tr(ai) $ & $ S (ai, ai+1) $ is the separation between an aircraft ai and its successor ai+1, and $ tr $ is the aircraft’s planned landing time. \\
		\hline Taxing & $ S (ai, ai+1) > tg (ai+1) – tg (ai) $& $ tg $ is the planned time for the entrance of the gateway for ai. \\
		\hline Parking & $ tk (ai+1) > tf (ai) $ & $ tk $ is the aircraft arriving time to a gate, and $ tf $ is its exit time. \\\hline 
	\end{tabular}
	\label{tab: sep}
\end{table}
\subsubsection{Maintaining aircraft planning qualities}
\begin{figure}[h!]
	\centering
	\includegraphics [scale = 0.58]{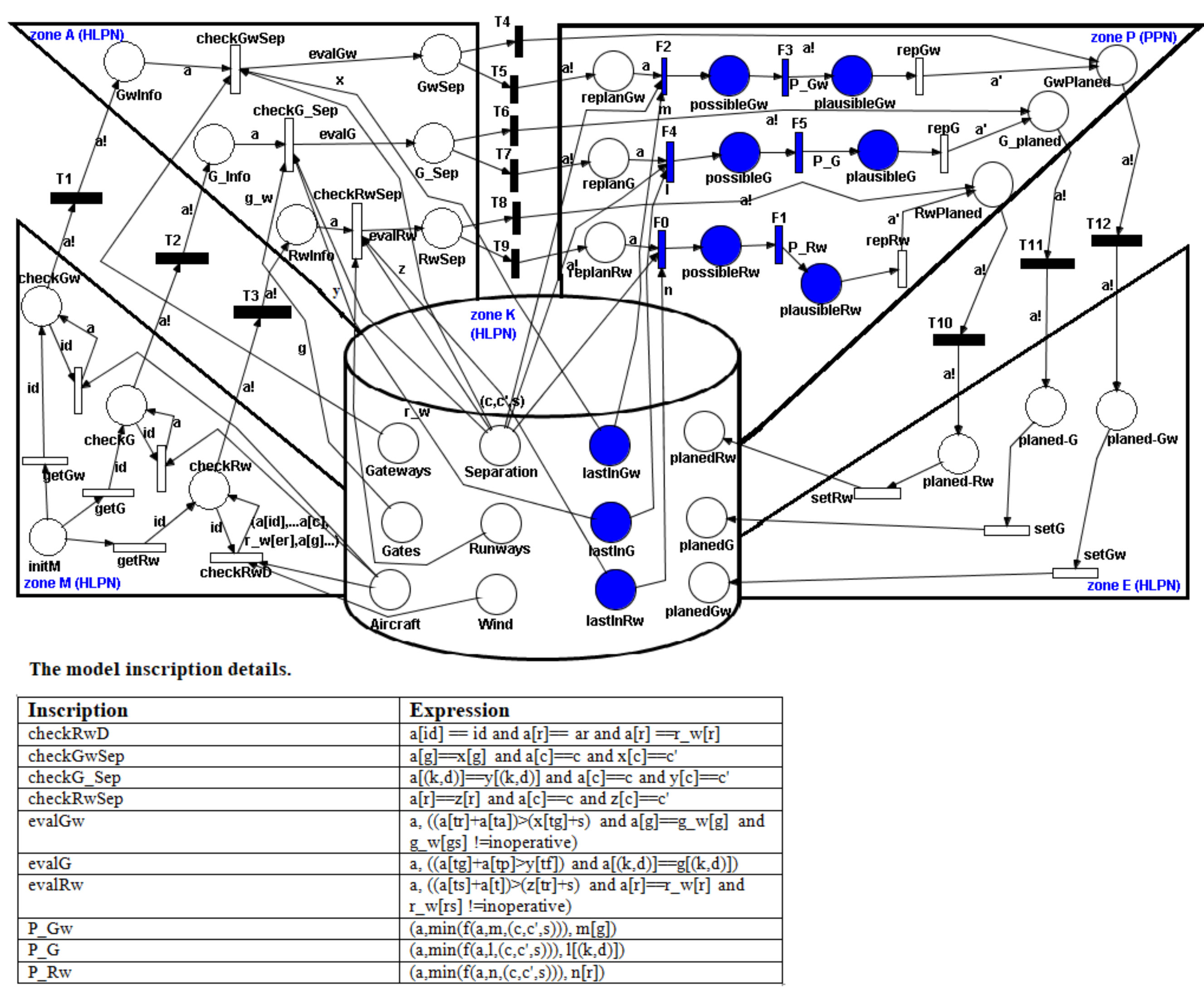}
	
	\caption{Maintaining safety and efficiency qualities of the Aircraft arrival procedure.}
	\label{fig: loop}
\end{figure}
With the aim of maintaining safety and efficiency constraints while adapting and updating the aircraft planning in the presence of a changing context, we define the model presented in Figure ~\ref{fig: loop}, where blue places and transitions are plausible and the black transitions (T1-T12) are the cross-zone transitions. Since the planning system is affected by several events its monitoring is required. Thus data to be monitored is: the wind direction, the airport resources states and the aircraft arrival time that allow determining the separation and the estimated overall time of the arrival the procedure.  

In the knowledge zone, several places are shared and used by both the managed and managing sub-systems such as: \textit{Aircraft} place, which contains the planned aircraft, \textit{lastInRw}, \textit{lastInGw} and \textit{lastInG} places; this data is used to calculate the separation distance between two consecutive aircraft in order to ensure the arrival procedure safety and the planning effectiveness. For the weather conditions, we considered only wind. The actual available resources of the airport (runways, gateways and gates) are represented by the places \textit{Runways}, \textit{Gateways} and \textit{Gates} respectively.

At the arrival procedure, the system uses the API read primitive \textit{getTokens} to obtain the actual parameters and observe context changes. This primitive is represented by the transitions \textit{getRw}, \textit{getGw} and \textit{getG}. Then, it checks the resource availability, if the aircraft is preparing for the landing phase, a step for checking wind direction is mandatory before landing. The analyzer firstly compares (tr(ai) + S (ai, ai+1)) with (ts(ai+1) +t(ai+1)) where ai is the last aircraft planned on the runway before the actual aircraft ai+1 and tr, S (ai, ai+1), ts and t represent the planned landing time, the required separation between the two aircraft, the time of the aircraft occurrence at the sequence point and the estimated aviation time to the landing point, respectively. 
The verification results are transferred to the planner where the PPN intervenes and the process of choosing a new resource is carried out. So, it calculates the possible release time of each runway by function f represented by the transition \textit{F0} of Figure ~\ref{fig: loop} where: $ f = tr + s;  $ 
with s is the separation between two consecutive aircraft obtained from the place Separation and tr is retrieved from the \textit{lastInRw} place, which consists of a vector containing data on the last landing for each runway. After calculating the release time of all runways, they are reduced to those checking the condition: $ ts + t > f . $  Only one runway will be then selected; the runway with the smallest value of f; this allows selecting the best runways and thus ensures the efficiency constraint. The selection is achieved using the transition \textit{F1} and its output arc expression, which selects the runway corresponding to the smallest f. The new runway identifier is assigned to the aircraft, and a new token is added to place \textit{planedRw}; the \textit{setTokens} write primitive of the API, represented by transition \textit{setRw}, is used and executed by the executor to achieve this adaptation.
\\	
Similar operations are performed for tokens of the \textit{Landed} place by firing transitions \textit{F2} and \textit{F3} to allocate a new gateway. For gate checking, tokens in the \textit{Taxied} place are recovered and used by the plausible transitions \textit{F4} and \textit{F5} to allocate a new gate to the aircraft in  case of the safety constraint violation.
\subsection{Aircraft planning validation}
\begin{figure}
	\centering
	\includegraphics [scale = 0.6]{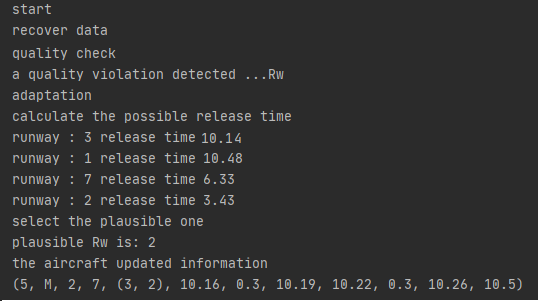}
	
	\caption{Simulation results of the self-adaptive aircraft arrival planning model.}
	\label{fig: sim}
\end{figure}
The proposed model is simulated and validated using the PNemu\footnote{PNemu has been released as open source software, available at https://github.com/SELab-unimi/pnemu.} framework. To achieve such validation, we have realized some extensions and updates on the PNemu, which are missing in the original version of the source code. More precisely, we defined a new class "HLPN" to model the managed system and its environment by an HLPN; we have modified the emulator class to be able to emulate models specified as HLPN; then we have redefined some primitives of the API to capture the HLPN concepts, such as \textit{getTokens} and \textit{setTokens} primitives.

For brevity, we show only one scenario corresponding to the runway re-planning; we also avoid defining all models in PNemu. We assume an initial configuration of the managed system presented in Figure ~\ref{fig: med}. The planned aircraft are presented in Table \ref{tab: sometab}.
\begin{table}
	\centering
	\caption{The actual arrival planning of the airport.}
	\begin{tabular}{|c|c|c|c|c|c|c|c|c|c|c|c|}
		\hline \textbf{Aircraft} & \textbf{c}& \textbf{r}& \textbf{g}& \textbf{(k, d)}& \textbf{ts}& \textbf{t}& \textbf{tr}& \textbf{tg}& \textbf{tp}& \textbf{tk}& \textbf{tf}  \\
		\hline Aircraft 1& L& 1& 7& (3, 2)& 10:15& 2& 10:18& 10:21& 2& 10:25& 10:45 \\
		\hline Aircraft 2& M& 3& 5& (1, 7)& 10:33& 3& 10:37& 10:40& 4& 10:45& 10:58 \\
		\hline Aircraft 3& H& 2& 9& (6, 8)& 14:15& 4& 14:18& 14:21& 1& 14:25& 14:45 \\
		\hline Aircraft 4& L& 7& 5& (1, 7)& 6:00& 2& 6:03& 6:07& 2& 6:10& 6:40 \\
		\hline \rowcolor{cyan} Aircraft 5&  M&  1&  7&  (3, 2)&  9:16&  3&  9:19&  9:22&  3&  9:26&  9:50 \\
		\hline Aircraft 6& H& 2& 9& (6, 8)& 16:05& 4& 16:10& 16:13& 2& 16:16& 16:40 \\
		\hline Aircraft 7& L& 3& 5& (3, 2)& 9:40& 3& 9:44& 9:48& 1& 9:50& 10:20 \\
		\hline Aircraft 8& M& 7& 7& (6, 8)& 14:21& 2& 14:24& 14:27& 2& 14:30& 14:45 \\
		\hline Aircraft 9& H& 2& 9& (1, 7)& 3:20& 2& 3:23& 3:27& 3& 3:31& 4:00 \\\hline 
	\end{tabular}
	\label{tab: sometab}
\end{table}
Place \textit{lastInRw} contains aircraft 1, 4, 7 and 9; places \textit{lastInGw} and \textit{lastInG} contain aircraft 9 and 7 respectively.
It is assumed that aircraft 5 is planned to arrive at 9:16h, but it is arriving at 10:16h. It finished the sequencing phase and it prepares for landing, so the runway verification process is started, at this time the planned runway is occupied by aircraft 1; runway re-planning is required.

The simulation results are as follows: the actual planning of aircraft 5 violates the safety constraints due to a delay in its arrival time, i.e. the separation distance is not maintained. To restore the separation constraint, the Planner element affects a new runway to aircraft 5, runway 2 for instance. A new token is deposited in the \textit{planedRw} place containing information (5, 2); the updated information for the aircraft is (5, 'M', 2, 7, (3, 2), 10.16, 0.3, 10.19, 10.22, 0.3, 10.26, 10.50). The adaptation details and the most plausible runway selection process are illustrated in Figure ~\ref{fig: sim}. If another aircraft is scheduled to land on that runway and arrives before the end of the landing operation of aircraft 5, it will be also re-planned.
\\
Since the model is a Petri net object, it is possible to compile it and use the compiled model along with the SPOT\footnote{[SPOT] https://spot.lrde.epita.fr} library to verify the correctness of the overall self-adaptive system with respect to design-time requirements expressed using LTL properties or another model checker.
\section{Conclusion}
This paper shows how to combine HLPNs and PPNs to model and analyze quality-driven self-adaptive systems evolving under uncertainty but still maintaining and guaranteeing the continuous satisfaction of an acceptable quality of service. HLPNs intervene in the definition of data flows through expression and annotation concepts, which are exploited to quantify the observed qualities among the different elements of the model. They are used for modelling the managed system and its execution context to improve the model expressiveness by representing more complex data structures of a dynamic system. PPNs are used to assist and improve the decision-making process in presence of uncertainty and hence determining the most appropriate adaptation plans through the concept of decision plausibility. An extension of the PNemu framework is also realized to take charge of HLPNs and PPNs introduced concepts as expressions and annotations. We evaluated the proposed model through the aircraft planning problem using the extended version of the PNemu framework. 

As future work, we intend to explore quantitative analysis techniques to predict the impact of an adaptation plan on the overall system quality to improve the decision-making process to deal with uncertainty in the selection of both the adaptation actions and side effects and the impact of the adaptation plan on system overall qualities. We also intend to combine machine learning techniques with Petri nets to better improve the decision-making and proactivity through model training and learning.
\bibliographystyle{eptcs}
\bibliography{biblio}
\end{document}